\documentclass[10pt,final,journal]{IEEEtran} 
\IEEEoverridecommandlockouts
\usepackage{amsmath,amssymb,amsfonts}

\usepackage[square, sort&compress, numbers]{natbib}

\usepackage[utf8]{inputenc}
\usepackage[T1]{fontenc}

\usepackage{pgfplots}
\pgfplotsset{compat=1.8}

\newtheorem{myrem}{Remark}

\DeclareMathOperator{\diag}{diag}
\DeclareMathOperator{\Expect}{\bf E}

\makeatletter
\newcommand*{\tsp}{%
	{\mathpalette\@tsp{}}%
}
\newcommand*{\@tsp}[2]{%
	\raisebox{\depth}{$\m@th#1\intercal$}%
}
\makeatother

\newcommand*\ie{\textit{i.e.}}
\newcommand*\eg{\textit{e.g.}}
\newcommand{\cond}{\,|\,}
\newcommand{\ith}{$ i $th }
\newcommand{\jth}{$ j $th }
\newcommand{\pD}{p_\textsc{d}}
\newcommand{\pS}{p_\textsc{s}}
\newcommand{\pDk}{p_{\textsc{d},k}}
\newcommand{\pSk}{p_{\textsc{s},k}}
\newcommand{\gaussian}[3]{\mathcal{N}(#1 \,;\, #2, #3)}
\newcommand{\gaussianbig}[3]{\mathcal{N}\bigl(#1 \,;\, #2, #3\bigr)}
\newcommand{\uniform}[2]{\mathcal{U}(#1 \,;\, #2)}

\begin{document}
	\bstctlcite{IEEEexample:BSTcontrol}
	
\title{PMBM filter with partially grid-based birth model with applications in sensor management
\thanks{This work was partially supported by the Wallenberg AI, Autonomous Systems and Software Program (WASP) funded by the Knut and Alice Wallenberg Foundation; the Industry Excellence Center LINKSIC funded by The Swedish Governmental Agency for Innovation Systems (VINNOVA); and Saab AB.}
}
\author{\IEEEauthorblockN{Per Boström-Rost, Daniel Axehill,~\IEEEmembership{Senior~Member,~IEEE}, Gustaf Hendeby,~\IEEEmembership{Senior~Member,~IEEE}}\\
\IEEEauthorblockA{\textit{Dept.\ of Electrical Engineering, Linköping University} \\
Linköping, Sweden}
}

\maketitle

\begin{abstract}
	This paper introduces a Poisson multi-Bernoulli mixture (PMBM) filter in which the intensities of target birth and undetected targets are grid-based. A simplified version of the Rao-Blackwellized point mass filter is used to predict the intensity of undetected targets, and the density of targets detected for the first time are approximated as Gaussian. Whereas conventional PMBM filter implementations typically use Gaussian mixtures to model the intensity of undetected targets, the proposed representation allows the intensity 
	to vary over the region of interest with sharp edges around the sensor's field of view, without using a large number of Gaussian mixture components. 
	This reduces the computational complexity compared to the conventional approach.
	The proposed method is illustrated in a sensor management setting where trajectories of sensors with limited fields of view are controlled to search for and track the targets in a region of interest.
\end{abstract}

\section{Introduction}
This paper considers multi-target tracking problems where the sensor's field of view is much smaller than the region of interest. 
In such scenarios, it is not only of interest to estimate the states of detected targets, but also to estimate where undetected targets may be located.
This can be done by combining information about areas that have been covered by the sensor with prior knowledge about where targets are likely to appear.
The problem is complicated as the number of targets is both unknown and time-varying, there are misdetections, false alarms, and unknown measurement origins.

The random finite set (RFS) framework enables a Bayesian approach to the considered problem \cite{mahler2014}. Several RFS-based methods to estimate the multi-target posterior density have been proposed. Examples of these include the probability hypothesis density (PHD) filter \cite{mahler2003}, the generalized labeled multi-Bernoulli (GLMB) filter \cite{vo2013,vo2014}, and the Poisson multi-Bernoulli mixture (PMBM) filter \cite{williams2015}.
The PMBM filter assumes that the multi-target density consists of a union of a Poisson point process (PPP) and a multi-Bernoulli mixture (MBM) \cite{williams2015}. The MBM component considers all possible data association hypotheses and can, \eg, be implemented using a track-oriented multiple hypotheses tracking (MHT) formulation \cite{garcia2018}. The PPP explicitly models the intensity of undetected targets, thereby indicating where it is likely to find new targets \cite{williams2015}.

In PMBM filter implementations, the birth intensity and intensity of undetected targets are commonly modeled as a single Gaussian or a Gaussian mixture \cite{garcia2018,frohle2019,bostrom2021}. While this approach is suitable when the sensor's field of view covers most of the region of interest, it is potentially inefficient if the field of view is small or if the birth intensity is uniform in the region of interest. 
Although a Gaussian mixture can approximate any intensity, a large number of components may be necessary to approximate a uniform density or to obtain sharp edges around regions that have recently been visited by the sensor. 
One alternative could be to apply the ideas of \cite{legrand2020}; start with a small number of components and partition the intensity along the boundaries of the field of view through recursive Gaussian splitting. However, this might still lead to a large number of components.
	
A different approach is proposed in \cite{beard2013}, which introduces a version of the Gaussian mixture PHD filter where it is assumed that new targets are always detected. Thereby, the complicated representation of undetected targets is avoided. Furthermore, by allowing the target birth intensity to take  on  a  uniform  distribution for the part of the target state that is directly observed through the measurements, the need for a large Gaussian mixture representation of target birth is eliminated \cite{beard2013}.
This approach can also be used in the PMBM filter but is not applicable in sensor management applications where the intensity of undetected targets is exploited when planning the search for new targets \cite{bostrom2020}.

This paper makes use of the ideas presented in \cite{beard2013} and proposes a version of the PMBM filter where the birth intensity is grid-based, \ie, a mixture of disjoint weighted uniform distributions, for the part of the target state that is observed through the measurements, and Gaussian distributed for the unobserved state components. 
The structure of the intensity is preserved in the prediction step, which is computed using a simplified version of the Rao-Blackwellized point mass filter \cite{smidl2013}.
The grid-based representation allows the intensity of undetected targets to vary over the region of interest with sharp edges around the sensor's field of view, without using a large number of Gaussian mixture components.
Furthermore, it is shown that the density of potential targets detected for the first time can be approximated as Gaussian.
This allows the prediction and measurement update of the MBM component to be performed as in a standard Gaussian PMBM filter implementation \cite{garcia2018}.

The proposed filter is compared to a conventional PMBM filter with Gaussian mixture birth in a sensor management application, where a team of sensors are monitoring an area in order to search for and track an unknown number of targets.

\section{Background}
This section provides an overview of random finite set (RFS) models and multi-target filtering using RFS theory.

\subsection{Random finite set models}
A \textit{random finite set} (RFS) is a set with a random number of elements which are themselves random \cite{mahler2007}. This means that the cardinality of an RFS is a discrete random variable, and its elements are random variables. This makes the RFS framework convenient for representing sets of multiple targets and multiple sensor measurements. In this work, the following RFSs will be of interest:
\subsubsection{Poisson point process}
A Poisson point process (PPP) is an RFS of which the cardinality is Poisson distributed with rate~$ \mu $ and the elements, given the cardinality, are independent and identically distributed (IID) according to $ p(x) $. The rate $ \mu $ and distribution $ p(x) $ form the intensity $ \lambda(x) $ of the PPP as $ \lambda(x) = \mu p(x). $ The density of a PPP $ X $ is given by \cite{ristic2013}:
\begin{equation}\label{eq:ppp_density}
\pi(X) = e^{-\langle \lambda , 1 \rangle}\prod_{x \in X} \lambda(x)\text{,}
\end{equation}
where the notation $ \langle a,b \rangle = \int a(x)b(x)\,\mathrm{d}x $ is used for the inner product of $ a(x) $ and $ b(x) $.

\subsubsection{Bernoulli RFS}
The cardinality of a Bernoulli RFS is Bernoulli distributed with parameter $ r \in [0,1] $. It is either empty, with probability~${ 1-r }$ or, with probability~$ r $, contains a single element with density~$ p(x) $. Thus, the density of a Bernoulli RFS $ X $ is \cite{ristic2013}:
\begin{equation} \label{eq:bernoulli_density}
\pi(X) = \begin{cases}
1-r, &X = \emptyset,\\rp(x), & X = \{x\}, \\ 0, &\vert X \vert \geq 2\text{.}
\end{cases}
\end{equation}

\subsubsection{Multi-Bernoulli RFS}
The disjoint union of a fixed number of Bernoulli RFSs is a multi-Bernoulli (MB) RFS. Its density is defined by the parameters $ { \{r^{i}, p^{i}  \}_{i\in \mathbb{I}} } $, where $ \mathbb{I} $ is an index set:
\begin{equation}\label{eq:mb_density}
\pi(X) = 
\begin{cases}
\sum_{\uplus_{i \in \mathbb{I}} X^i = X} \prod_{i \in \mathbb{I}} \pi^{i}(X^i), &\vert X \vert \leq \vert \mathbb{I} \vert, \\
0, & \vert X \vert  > \vert \mathbb{I} \vert.
\end{cases}
\end{equation}
The notation $ X^1 \uplus X^2=X $ denotes disjoint union, \ie, $ {X^1 \cup X2=X} $ and $ {X^1 \cap X^2=\emptyset}$.

\subsubsection{Multi-Bernoulli mixture RFS}
A normalized, weighted sum of MB RFSs is referred to as a multi-Bernoulli mixture (MBM) RFS. 
Its density can be expressed as \cite{granstrom2020}:
\begin{equation}\label{eq:mbm_density}
\pi(X) = \sum_{j \in \mathbb{J}} w^j \sum_{\uplus_{i \in \mathbb{I}^j} X^i = X}  \prod_{i \in \mathbb{I}^j} \pi^{j,i}(X^i)
\end{equation}
and is defined by the set of parameters $ { \{w^j, \{r^{j,i}, p^{j,i} \}_{i\in \mathbb{I}^j} \}_{j \in \mathbb{J}} } $, where $ \mathbb{J} $ is an index set for the MB components of the MBM, $ \mathbb{I}^j $ is an index set for the Bernoullis of the $ j $th MB RFS, and $ w^j $ is the weight of the $ j $th MB.

\subsection{Multi-target filtering using RFS theory}\label{sec:rfs_filtering}
In the RFS-based filtering approach, the multi-target state and set of measurements at time $ k $ are modeled as two RFSs denoted $ X_k $ and $ Z_k $, respectively. The aim is to estimate the posterior multi-target state density $ \pi_{k \mid k}(X_k \cond Z_{1:k}) $, where $ Z_{1:k} $ is a collection of finite sets of measurements received up to time $ k $.

Similar to the standard single-target case, the multi-target posterior density can be computed recursively via prediction and measurement update steps. With the Bayes multi-target filter \cite{mahler2007}, the posterior multi-target density at time $ k $ is propagated in time using the  Chapman-Kolmogorov equation
\begin{equation}\label{eq:mttbayes_prediction}
\pi_{k+1 \mid k} (X \cond Z_{1:k}) =
\int \phi_X (X \cond X') \pi_{k \mid k}(X' \cond Z_{1:k})\, \delta X', 
\end{equation}
where $ \phi_X (X \cond X') $ is the standard multi-target transition density, \ie, a Markovian process for individual targets with transition density $ p_{k+1,k}(x\cond x') $ and state-dependent probability of survival $ p_{\textsc{s},k}(x) $, combined with a PPP birth process with intensity $ \lambda^\text{b}_k(x) $.

Given a measurement set $ Z_k $ with multi-target measurement set likelihood function $ \phi_Z(Z_k \cond X_k) $, the predicted multi-target density is updated using Bayes' rule
\begin{equation}\label{eq:mttbayes_update}
\pi_{k \mid k} (X \cond Z_{1:k}) = \frac{\phi_Z(Z_k \cond X) \pi_{k \mid k-1} (X \cond Z_{1:k-1})}{\int \phi_Z(Z_k \cond X) \pi_{k \mid k-1} (X \cond Z_{1:k-1})\, \delta X}.
\end{equation}
The standard multi-target  measurement  model \cite{mahler2007} is used, which means that $ \phi_Z(Z_k \cond X_k) $ models  noisy measurements of individual targets with state-dependent probability of detection $ p_{\textsc{d},k}(x) $, combined with PPP clutter with intensity $ \lambda^\text{fa}_k(z) $. At most one measurement is generated for each target in each time step, and each measurement is the result of at most one target. A measurement of a target is independent of all other targets and other measurements conditioned on the same target; the single target measurement likelihood is $ p_{z_k}(x) = p(z_k \cond x) $. Note that the above functions in general depend on the sensor state, but this is implicit here for notational brevity. The integrals in \eqref{eq:mttbayes_prediction} and \eqref{eq:mttbayes_update} are set integrals, as defined in \cite{mahler2007}.

\section{PMBM filter with partially uniform target birth model}
The PMBM filter estimates the state of the set of targets, \ie, the density $ \pi $ of the RFS $ X $, which is assumed to be a PMBM:~a union of a PPP component~\eqref{eq:ppp_density} and an MBM component~\eqref{eq:mbm_density}. The PPP represents targets that are hypothesized to exist, but have never been detected, \eg, targets that have been located in a region where the sensor system has low detection probability. The MBM represents targets that have been detected at least once, and each MB in the mixture corresponds to a unique sequence of data associations for all detected targets, referred to as a global hypothesis.


The general form of the PMBM filter recursion for the intensity of undetected targets $ \lambda^\text{u}(x) $ is given by
\begin{subequations}\label{eq:pmbm_ppp_recursion}
	\begin{align}
	\lambda^\text{u}_{k+1 \mid k}(x) &= \lambda^\text{b}_{k+1}(x) \nonumber \\&\qquad+ \int \pSk (x') p_{k+1,k}(x \cond x') \lambda^\text{u}_{k \mid k}(x') \,\text{d}x' \text{,} \label{eq:pmbm_ppp_prediction} \\
	\lambda^\text{u}_{k \mid k}(x) &= (1-\pDk(x))		\lambda^\text{u}_{k \mid k-1}(x) \text{,} \label{eq:pmbm_ppp_update}
	\end{align}
\end{subequations}
where $ \lambda^\text{b}_k(x) $ is the intensity of target birth at time $ k $, $ \pSk(x) $ is the probability of survival for a target in state $ x $ at time $ k $, $ p_{k+1,k}(x \cond x') $ is the target state transition density from time $ k $ to time $ k+1 $, and $ \pDk(x) $ is the probability of detection for a target in state $ x $ at time $ k $.

Each measurement $ z_k $ at time $ k $ generates a new potentially detected target, which is represented by a Bernoulli RFS with existence probability $ r(z_k) $ and density $ p(x \cond z_k) $ defined as
\begin{subequations}\label{eq:pmbm_new_bernoulli}
	\begin{align}
	r(z_k) &= \frac{\int \pDk(x) p_{z_k}(x) \lambda^\text{u}_{k \mid k-1}(x) \, \text{d}x}{\lambda^\text{fa}(z_k) + \int \pDk(x) p_{z_k}(x) \lambda^\text{u}_{k \mid k-1}(x) \, \text{d}x} \text{,} \label{eq:pmbm_new_bernoulli_existence_probability}\\
	p(x \cond z_k) &= \frac{\pDk(x) p_{z_k}(x)  \lambda^\text{u}_{k \mid k-1}(x)}{\int \pDk(x) p_{z_k}(x) \lambda^\text{u}_{k \mid k-1}(x) \, \text{d}x}\text{,} \label{eq:pmbm_new_bernoulli_density}
	\end{align}
\end{subequations}
where $ p_{z_k}(x) = p(z_k \cond x) $ is the measurement likelihood function for a measurement $ z_k $ given target state $ x $.

In the following sections, approximations of \eqref{eq:pmbm_ppp_recursion} and \eqref{eq:pmbm_new_bernoulli} are derived for the case when the birth intensity is a union of a grid-based intensity and a Gaussian component.
For the prediction and measurement update of existing tracks, the equations of the standard Gaussian implementation of the PMBM filter \cite{garcia2018} can be used. The equations corresponding to these steps are outlined in Appendix~\ref{sec:pmbm_filter_recursion}.

\subsection{Modeling}
Using the notation of \cite{beard2013}, let $ \theta $ represent the scalar or vector consisting of the part of the state $ x $ that directly affects the measurement likelihood, \ie, $ p_z(x) = p_z(\theta) $, and let $ \varphi $ represent the remaining part of $ x $.
In the following, $ x $ and $ (\theta, \varphi) $ are used interchangeably.

The intensity of undetected targets is modeled as the product of a grid-based intensity in~$ \theta $, \ie, a mixture of weighted disjoint uniform distributions, and a Gaussian distribution in~$ \varphi $. With $ {k' \in \{ k,k-1  \}} $ it can be expressed as
\begin{equation}\label{eq:intensity_of_undetected_targets_desired_form}
\lambda^\text{u}_{k \mid k'}(\theta, \varphi) = 
\biggl[\sum_{i=1}^N w^{(i)}_{k \mid k'} \uniform{\theta}{\mathcal{C}^{(i)}}
\biggr]
\gaussian{\varphi}{\hat{\varphi}}{P^\varphi}
\end{equation}
where $ \uniform{\theta}{\mathcal{C}^{(i)}} $ is the uniform distribution in~$ \theta $ over the cell~$ \mathcal{C}^{(i)} $, $ \hat{\varphi} $~is the prior mean of the unmeasured state component and $ P^\varphi $ is its variance, and $ w^{(i)}_{k \mid k'} $~is the expected number of undetected targets with state component $ \theta $~in cell $ \mathcal{C}^{(i)} $. 
The cells are assumed to be mutually disjoint, \ie, $ \mathcal{C}^{(i)} \cap \mathcal{C}^{(j)} = \emptyset $ for $ i \neq j $, and the union of all cells $ \bigcup_i \mathcal{C}^{(i)} $ covers the entire tracking volume in $ \theta $. The midpoint of cell $ \mathcal{C}^{(i)} $ is denoted $ \theta^{(i)} $.

For the typical case where the target state consists of position and velocity, and the measurements correspond to the position, the model \eqref{eq:intensity_of_undetected_targets_desired_form} corresponds to using a single prior distribution on the velocity in the entire region of interest and allowing the intensity of undetected targets be position-dependent.

It is assumed that both the probability of survival $ \pSk(x) $ and the probability of detection $ \pDk(x) $ are independent of $ \varphi $ and constant in each cell, \ie, $ \pSk(x) = \pSk^{(i)} $ and $ \pDk(x) = \pDk^{(i)} $ for all $ x $  such that $ \theta \in \mathcal{C}^{(i)} $. Furthermore, the measurement likelihood is assumed independent of $ \varphi $ and Gaussian distributed according to $ p_z(x) = p_z(\theta) = \gaussian{z}{H_k \theta }{R_k} $, where $ H_k $ is assumed invertible. Note that nonlinear measurement models can be handled using linearization (cf.\ extended Kalman filter (EKF)). A non-invertible $ H_k $ is more problematic but would in practice be handled by introducing additional prior information.  The target state transition density is linear Gaussian, \ie, $ p_{k+1,k} (x \cond x') = \gaussian{x}{F_k x'}{Q_k} $. Let the components of $ x $ be ordered such that the state transition function has the following structure:
\begin{equation}
	\underbrace{\begin{bmatrix}\theta_{k+1} \\ \varphi_{k+1} \end{bmatrix}}_{x_{k+1}} =
	\underbrace{\begin{bmatrix}F^\theta_k & F^{\theta\varphi}_k \\ F^\varphi_k & F^{\varphi\theta}_k \end{bmatrix}}_{F_k} 
	\underbrace{\begin{bmatrix}\theta_{k} \\ \varphi_{k} \end{bmatrix}}_{x_k} +
	\underbrace{\begin{bmatrix}w^\theta_{k} \\ w^\varphi_{k} \end{bmatrix}}_{w_k}
\end{equation}
where $ p(w_k) = \gaussian{w_k}{0}{Q_k} $. For simplicity, without loss of generality, the matrix $ Q_k $ is assumed to be block-diagonal, \ie, 
\begin{equation}
	Q_k =
	\begin{bmatrix} Q_k^\theta & Q_k^{\theta\varphi} \\ (Q_k^{\theta\varphi})^\tsp & Q_k^\varphi \end{bmatrix} = 
	\begin{bmatrix} Q_k^\theta & 0 \\ 0 & Q_k^\varphi \end{bmatrix}.
\end{equation}
If $ Q_k^{\theta\varphi} \neq 0 $, the state transition function can be transformed to ensure that the process noises acting on $ \theta $ and $ \varphi $ are independent \cite{schon2005}.

\subsection{Prediction}
Let the posterior intensity of undetected targets at time $ k $ be defined as in \eqref{eq:intensity_of_undetected_targets_desired_form} with $ k' = k $
and define the birth intensity as
\begin{equation}\label{eq:ppp_prediction_approximation:birth_intensity}
\lambda^\text{b}_k(\theta, \varphi) = 
\biggl[\sum_{i=1}^N w^{\text{b}(i)}_k \uniform{\theta}{\mathcal{C}^{(i)}}
\biggr]
\gaussian{\varphi}{\hat{\varphi}}{P^{\varphi}}  \text{,}
\end{equation}
where $ w^{\text{b}(i)}_k $ is the expected number of targets with state component $ \theta $ in cell $ \mathcal{C}^{(i)} $ appearing at time $ k $.
The aim is to find an approximation of \eqref{eq:pmbm_ppp_prediction} such that the predicted intensity $ \lambda^\text{u}_{k+1 \mid k} $ has the same form as \eqref{eq:intensity_of_undetected_targets_desired_form}. The birth intensity \eqref{eq:ppp_prediction_approximation:birth_intensity}, which represents the first term in \eqref{eq:pmbm_ppp_prediction}, already has the same form as \eqref{eq:intensity_of_undetected_targets_desired_form}. This means that only the second term (the integral) in \eqref{eq:pmbm_ppp_prediction} needs to be approximated such that it does not affect the intensity in $ \varphi $ and maintains the grid-based intensity in~$ \theta $.

The Rao-Blackwellized point mass filter (RB-PMF) \cite{smidl2013} is used to approximate the predicted intensity. Given a posterior point-mass density
\begin{equation}
	\hat{p}_{k \mid k}(x) = \sum_{i=1}^{N} w_{k \mid k}^{(i)}  \uniform{\theta}{\mathcal{C}^{(i)}} \gaussian{\varphi}{\hat{\varphi}^{(i)}_{k \mid k}}{P^{\varphi,(i)}_{k \mid k}} \text{,}
\end{equation}
the RB-PMF prediction gives predicted weights $ w_{k+1 \mid k}^{(j)} $ and a Gaussian distribution in $ \varphi_{k+1} $ for each combination of $ \theta^{(i)}_{k} $ and $ \theta^{(j)}_{k+1} $. This results in a predictive point-mass density \cite{smidl2013,dunik2019}
\begin{align}
&\hat{p}_{k+1 \mid k}(x) \nonumber \\
&\ \ = \sum_{j=1}^{N} \sum_{i=1}^{N} w_{k+1 \mid k}^{(i,j)} \uniform{\theta}{\mathcal{C}^{(j)}}
\gaussian{\varphi}{\hat{\varphi}_{k+1 \mid k}^{(i,j)}}{P^{\varphi,(i,j)}_{k+1 \mid k}} \text{,}
\end{align}
where
\begin{subequations}\label{eq:rbpmf_nonlinear_prediction}
	\begin{align}
	w_{k+1 \mid k}^{(i,j)} &= w_{k \mid k}^{(i)} \hat{\tilde{c}}^{(i,j)} \\
	y_{k+1}^{(i,j)} &= \theta_{k+1}^{(j)} - F^\theta_k \theta_k^{(i)}  \\
	\hat{\tilde{c}}^{(i,j)} &= \gaussian{y_{k+1}^{(i,j)}}
	{F^{\theta\varphi}_k \hat{\varphi}_{k \mid k}^{(i)}}
	{F^{\theta\varphi}_k P^{\varphi,(i)}_{k \mid k} (F^{\theta\varphi}_k)^\tsp + Q^{\theta}_k}
	\end{align}
\end{subequations}
and 
\begin{subequations}\label{eq:rbpmf_linear_prediction}
	\begin{align}
	\hat{\varphi}_{k+1}^{(i,j)} &= \Expect \bigl[ \varphi_{k+1} \cond \theta_k = \theta^{(i)}, \theta_{k+1} = \theta^{(j)}, z_{1:k}	\bigr] \nonumber \\
	&= F^{\varphi\theta}_k \theta^{(i)} + F^{\varphi}_k \hat{\varphi}^{(i,j)}_{k \mid k} \\
	P_{\varphi,k+1}^{(i,j)} &= \text{cov} \bigl[ \varphi_{k+1} \cond \theta_k = \theta^{(i)}, \theta_{k+1} = \theta^{(j)}, z_{1:k}	\bigr] \nonumber \\
	&= F^{\varphi}_k P_{\varphi,k \mid k}^{(i,j)} (F^{\varphi}_k)^\tsp + Q^{\varphi}_k \\
	\hat{\varphi}_{k \mid k}^{(i,j)} &= \hat{\varphi}_{k \mid k}^{(i)} + K_k^{(i,j)} (y_{k+1}^{(i,j)} - F^{\theta\varphi}_k \hat{\varphi}_{k \mid k}^{(i)})
	\\
	P^{\varphi,(i,j)}_{k \mid k} &= P^{\varphi,(i)}_{k \mid k} - K_k^{(i,j)} F^{\theta\varphi}_k P^{\varphi,(i)}_{k \mid k} \\
	K_k^{(i,j)} &= P^{\varphi,(i)}_{k \mid k} (F^{\theta\varphi}_k)^\tsp \bigl(F^{\theta\varphi}_k P^{\varphi,(i)}_{k \mid k} (F^{\theta\varphi}_k)^\tsp + Q^{\theta}_k\bigr)^{-1}
	\end{align}
\end{subequations}

The increase in the number of Gaussian components means that the complexity of the RB-PMF is exponential, and the typical remedy to this is to merge the Gaussian mixture in $ \varphi_{k+1} $ at each grid point $ \theta_{k+1}^{(j)} $ using moment matching \cite{smidl2013}.
Here, the mixtures are instead replaced with the single Gaussian distribution $ \gaussian{\varphi}{\hat{\varphi}}{P^{\varphi}} $. 
This means that no information about the state components $ \varphi $ of undetected targets is inferred from the varying intensity in the state components~$ \theta $. Instead, the same prior on $ \varphi $ is used in all grid points.
As a consequence of the simplification, \eqref{eq:rbpmf_linear_prediction} does not need to be computed and the following approximation of the prediction step for the intensity of undetected targets \eqref{eq:pmbm_ppp_prediction} is obtained:
\begin{equation}\label{eq:pmbm_ppp_prediction_approx}
\lambda^\text{u}_{k+1 \mid k}(\theta, \varphi) \approx 
\biggl[\sum_{j=1}^N w^{(j)}_{k+1 \mid k} \uniform{\theta}{\mathcal{C}^{(j)}}
\biggr]
\gaussian{\varphi}{\hat{\varphi}}{P^\varphi}
\end{equation}
where
\begin{equation}
w^{(j)}_{k+1 \mid k} = 
w^{\text{b}(j)} + \sum_{i=1}^N \pSk^{(i)} w_{k+1 \mid k}^{(i,j)} 
\end{equation}
and $ w_{k+1 \mid k}^{(i,j)}  $ is given by \eqref{eq:rbpmf_nonlinear_prediction} with $ \hat{\varphi}_{k \mid k}^{(i)} = \hat{\varphi} $ and $ P^{\varphi,(i)}_{k \mid k} = P^\varphi $. 
Note that \eqref{eq:pmbm_ppp_prediction_approx} is in the same form as \eqref{eq:pmbm_ppp_prediction} and that the prediction step computationally corresponds to a multi-dimensional convolution.

\begin{myrem}
	Note that if the computational complexity is not an issue, the full RB-PMF could be used to represent the intensity of undetected targets. However, this approach quickly becomes computationally expensive for anything but small examples.
\end{myrem}

\begin{myrem}
	The approach can also be extended to handle birth intensities with several Gaussian components in $ \varphi $, each with an associated grid-based intensity in $ \theta $. For the typical case where the target state consists of position and velocity and the measurements correspond to the position, this could be used to model that targets are expected to appear with different velocities in different regions of the tracking volume.
\end{myrem}

\subsection{Update}
Let the predicted intensity of undetected targets at time $ k $ be in the form \eqref{eq:intensity_of_undetected_targets_desired_form}, with $ k' = k-1 $.

\subsubsection{Intensity of undetected targets} 
The updated intensity of undetected targets \eqref{eq:pmbm_ppp_update} is straightforward to compute as
\begin{equation}
\lambda^\text{u}_{k \mid k}(\theta, \varphi)=
\sum_{i=1}^N w_{k \mid k}^{(i)} \uniform{\theta}{\mathcal{C}^{(i)}} \gaussian{\varphi}{\hat{\varphi}}{P^\varphi} \text{,} 
\end{equation}
where $ w_{k \mid k}^{(i)} = (1-\pDk^{(i)}) w_{k \mid k-1}^{(i)}$.
\subsubsection{Potential targets detected for the first time} 
The aim here is to find an approximation of \eqref{eq:pmbm_new_bernoulli} such that $ p (x \cond z_k ) $ is a Gaussian distribution. This is done by adapting the approach of~\cite{beard2013}, where the intensity in $ \theta $ is assumed uniform in the entire region of interest. Note that the update step presented here collapses to the one in \cite{beard2013} when a single cell $ \mathcal{C}^{(i)} $ is used, \ie, $ N=1 $.

In a first step, a set of cell indices $ \mathbb{I}_{z_k} $ is selected such that the probability mass of $ p_{z_k} (\theta) $ outside the supercell
\begin{equation}
	\mathcal{C}_{z_k} = \bigcup_{i \in \mathbb{I}_{z_k}} \mathcal{C}^{(i)}
\end{equation}
is negligible, \ie,
\begin{equation}\label{eq:likelihood_negligible}
	\int_{\mathcal{C}_{z_k}} p(z_k \cond \theta)\,\text{d}\theta \approx \int p(z_k \cond \theta) \,\text{d}\theta .
\end{equation}
Then, to make use of the approach of \cite{beard2013}, the intensity in $ \theta $ within $ \mathcal{C}_{z_k} $ is approximated as uniformly distributed, which means that
\begin{equation}\label{eq:supercell_approximation1}
	\sum_{i\in\mathbb{I}_{z_k}} \pDk^{(i)} w_{k \mid k-1}^{(i)} \uniform{\theta}{\mathcal{C}^{(i)}}  
	\approx 
	\uniform{\theta}{\mathcal{C}_{z_k}}  \sum_{i\in\mathbb{I}_{z_k}} \pDk^{(i)} w_{k \mid k-1}^{(i)} .
\end{equation}	
Using \eqref{eq:likelihood_negligible} and \eqref{eq:supercell_approximation1}, the product $ \pDk(x) p_{z_k}(x) \lambda^\text{u}_{k \mid k-1}(x) $ which is used in \eqref{eq:pmbm_new_bernoulli} can be written as
\begin{align}
&\pDk(x) p_{z_k}(x) \lambda^\text{u}_{k \mid k-1}(x) \nonumber 
\\ &\qquad=
p_{z_k}(\theta) \sum_{i=1}^N \pDk^{(i)} w_{k \mid k-1}^{(i)} \uniform{\theta}{\mathcal{C}^{(i)}} \gaussian{\varphi}{\hat{\varphi}}{P^\varphi}
\nonumber 
\\ &\qquad\approx
\sum_{i\in\mathbb{I}_{z_k}} \pDk^{(i)} w_{k \mid k-1}^{(i)} / \vert \mathcal{C}^{(i)} \vert 
p_{z_k}(\theta)
\gaussian{\varphi}{\hat{\varphi}}{P^\varphi}. \label{eq:pmbm_new_bernoulli_product_approx1}
\end{align}

Furthermore, using Theorem~2.1 in Chapter~1 of \cite{gut1995}, the Gaussian measurement likelihood implies that
\begin{equation}\label{eq:transformed_measurement}
	p_{z_k}(\theta) = \gaussian{z_k}{H_k}{R} = \det H_k^{-1} \gaussianbig{\theta}{\hat{\theta}(z_k)}{P_\theta(z_k)} \text{,}
\end{equation}
where
\begin{subequations}
	\begin{align}
	\hat{\theta}(z_k) &= H_k^{-1} z_k \text{,} \\
	P_\theta(z_k) &= H_k^{-1} R_k (H_k^{-1})^\tsp.
	\end{align}
\end{subequations}

Using \eqref{eq:pmbm_new_bernoulli_product_approx1} and \eqref{eq:transformed_measurement}, the desired approximation of \eqref{eq:pmbm_new_bernoulli} is obtained as
\begin{subequations}\label{eq:pmbm_new_bernoulli_approx}
	\begin{align}
	r(z_k) &\approx 
	\frac{ 
		\sum_{i\in\mathbb{I}_{z_k}} \pDk^{(i)} w_{k \mid k-1}^{(i)}  / \vert \mathcal{C}^{(i)} \vert \det H_k^{-1}}
		{\lambda^\text{fa}(z_k) + \sum_{i\in\mathbb{I}_{z_k}} \pDk^{(i)} w_{k \mid k-1}^{(i)}  / \vert \mathcal{C}^{(i)} \vert \det H_k^{-1}} \text{,} \label{eq:pmbm_new_bernoulli_approx_existence_probability}\\
	p(x \cond z_k) &\approx 
	\gaussianbig{\theta}{\hat{\theta}(z_k)}{P_\theta(z_k)}
	\gaussian{\varphi}{\hat{\varphi}}{P^\varphi}\text{.} \label{eq:pmbm_new_bernoulli_approx_density}
	\end{align}
\end{subequations}

\section{Application to sensor management}
In this section, the proposed version of the PMBM filter is used as an underlying estimator in a multi-target sensor management problem. In the considered scenario, a team of controllable sensors are used to search for and estimate the states of an unknown number of targets. 
As the PMBM filter not only provides estimated states of the discovered targets, but also a representation of where previously undetected targets are likely to be found, it provides an appealing foundation for a unified search and track method \cite{bostrom2020}.
Based on the PMBM density, the sensors' control inputs can be optimized in order to simultaneously maintain track of discovered targets and search for targets that are yet to be detected.

\subsection{Problem formulation}
At time $ k $, the problem of computing the control inputs for a team of $ S $ sensors for time steps $ k $ up to $ k+T-1 $ is formulated as the following stochastic optimal control problem,
\begin{equation}\label{eq:problem_formulation}
\begin{split}
\begin{alignedat}{3}
& \underset{u_k^{1:S}}{\text{minimize}} \quad	&& \Expect \Biggl[ \sum_{t=k+1}^{k+T} \ell(\pi_{t \mid t}) \Biggr]  \\
& \text{subject\ to} 					&& u_k^j \in \mathcal{P}_k^j, \\
& 										&& s_t^j = g_t(s_k^j, u_k^j), \\
& 										&& \pi_{t+1 \mid t+1} = \rho(\pi_{t \mid t} , \hat{Z}_{t+1}(u_k^{1:S}), s_{t+1}^{1:S} \bigr)\text{,}\\
\end{alignedat}
\end{split}
\end{equation}
where the expectation is taken with respect to the future measurement sets and
\begin{itemize}
	\item $ \pi_{t \mid t} $ is the posterior PMBM density at time $ t $,
	\item $ T $ is the planning horizon, 
	\item $ u_k^j $ is a sequence of control inputs from times $ k $ up to $ k+T-1 $ for sensor $ j $,
	\item $ \mathcal{P}_k^j $ is a discrete set of all admissible sequences of control inputs at time $ k $ for sensor $ j $,
	\item $ g_t(s_k,u_k) $ is a function that returns the sensor state at time $ t $ if control sequence $ u_k $ is applied to a sensor with state $ s_k $ at time $ k $,
	\item $ u_k^{1:S} $ is a collection of control sequences for $ S $ sensors,
	\item $ \hat{Z}_t(u_k^{1:S}) $ is a hypothesized measurement set at time $ t $, assuming that control input sequences $ u_k^{1:S} $ were selected at time $ k $,
	\item $ s_t^{1:S} $ is a collection of sensor states at time $ t $,
	\item $ \rho(\cdot) $ is a shorthand notation for a prediction step followed by an update step in the PMBM filter, and
	\item $ \ell(\cdot) $ is a stage cost function used to trade-off between tracking discovered targets and searching for new targets.
\end{itemize}
Since the targets maneuver and the scenario changes over time, it is necessary to re-plan the sequences of sensor control inputs online as new measurements are obtained. 
This is done in standard receding horizon fashion~\cite{maciejowski2002}, \ie, for each sensor, only the first element of the computed sequence of control inputs is applied before re-planning.

\subsubsection{Predicted measurement sets}
The computationally expensive expectation over the future measurement sets in \eqref{eq:problem_formulation} is avoided using the predicted ideal measurement set approach \cite{mahler2004}. This corresponds to assuming that the measurement sets $ \hat{Z}_{k+1:k+T}(u_k^{1:S}) $ are generated without misdetections, measurement noise, and false alarms.

\subsubsection{Objective function}
The objective function proposed in \cite{bostrom2020} is employed, \ie, the cost function $ \ell(\pi_{t \mid t}) $ is a weighted sum of two terms, 
\begin{equation}\label{eq:proposed_Cost_function}
\ell(\pi_{t \mid t}) = \ell^\text{track} (\pi_{t \mid t}) + \eta \ell^\text{search}(\pi_{t \mid t}),
\end{equation}
where $ \ell^\text{track} $ captures the tracking performance, $ \ell^\text{search} $ captures the search performance, and $ \eta $ is a user-defined weight used to trade-off between tracking and searching \cite{bostrom2020}.

\subsubsection{Greedy assignment}
The optimization problem in \eqref{eq:problem_formulation} scales exponentially with the number of sensors. Instead of solving the problem jointly for all sensors, the sequentially greedy approximation strategy of \cite{dames2015} is utilized to assign control sequences to each sensor. Initially, each sensor computes the objective function value for each control sequence $ u^j_k \in \mathcal{P}^j_k $, without considering the other sensors. The combination of sensor and control sequence with the smallest objective function value is selected. The remainder of the team then recomputes the objective function value of each control sequence conditioned on the selected control sequence of the first sensor, and the sensor and control sequence with the lowest cost are again selected. This process repeats until all sensors have been assigned a control sequence.

\subsection{Simulation results}
\begin{figure}
	\centering
	\includegraphics{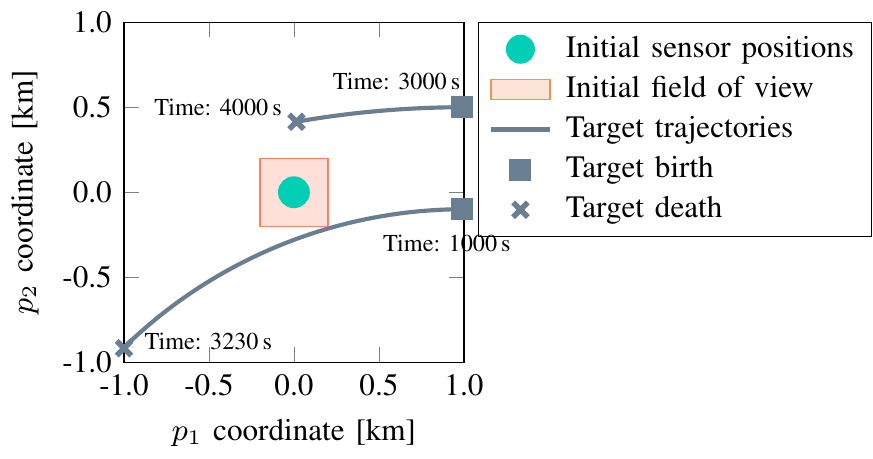}
	\caption{Target trajectories and initial sensor states in the considered scenario.}
	\label{fig:scenario_geometry}
\end{figure}

The simulated scenario involves a team of two sensors with limited fields of view that are tracking an unknown and time-varying number of targets.
No targets are present at the beginning of the considered scenario, with one arriving at time 1000\,s, and one more at time 3000\,s. The first target leaves the scene at 3230\,s and the second one remains until the end of the scenario, which lasts for 4000\,s. Fig.~\ref{fig:scenario_geometry} illustrates the scenario geometry.

The single-target state $ x = [p_1,v_1,p_2,v_2]^\tsp $ consists of its two-dimensional position and velocity and its dynamics follow a nearly constant velocity model, \ie, $ p_{k+1,k}(x \cond x') = \gaussian{x}{F_\textsc{cv} x'}{Q} $, where 
\begin{equation}
F_\textsc{cv} = 
I_2 \otimes
\begin{bmatrix}
1 & \tau \\ 0 & 1
\end{bmatrix},
\ \ \ 
Q = 
\sigma_w^2 GG^\tsp,
\ \ \ 
G= I_2 \otimes 
\begin{bmatrix}
\frac{\tau^2}{2} \\ \tau
\end{bmatrix},
\end{equation}
where $ \otimes $ is the Kronecker product, $ \tau= 10\,\text{s} $ is the sampling period, and $ \sigma_w = 0.05\,\text{m}/\text{s}^2 $ is the standard deviation for the target acceleration. The survival probability is assumed constant $ p_\textsc{s} = 0.99 $ for each target.

A sensor with state $ s $ at position $ (p_1^\text{s},p_2^\text{s}) $ has field of view $ \mathcal{V}(s) $ in the form of a square with sides $ a=400\,\text{m} $ centered around its position,
\begin{equation}
\mathcal{V}(s) = \{ x \,:\, \max\bigl( \vert p_1 - p_1^s \vert, \vert p_2 - p_2^s\vert \bigr) \leq a/2 \}.
\end{equation}
The detection probability is constant within the field of view and each detection results in a linear measurement of the position of the corresponding target, \ie,
\begin{subequations}
\begin{align}
\pD(x \cond 	s) &= 
\begin{cases}
0.9, &\text{if } x \in \mathcal{V}(s) \text{,}\\
0, &\text{otherwise.}
\end{cases}
\\
p(z \cond x,s) &= 
\begin{cases}
\gaussian{z}{Hx}{R}, &\text{if } x \in \mathcal{V}(s) \text{,}\\
0, &\text{otherwise.}
\end{cases}
\end{align}
\end{subequations}
with parameters
\begin{equation}
H=I_2\otimes
\begin{bmatrix}
1 & 0
\end{bmatrix}
\text{,}
\quad
R = \sigma_p^2 I_2.
\end{equation}
where $ \sigma_p = 10\,\text{m} $ is the standard deviation of the position measurement noise.
The clutter is modeled by a Poisson RFS with uniform intensity in the field of view and five expected false alarms per time step.

The part of the target state $ x $ that is involved in the measurement $ z $ corresponds to the position components, \ie, $ \theta = [p_1,p_2]^\tsp $ and $ \varphi = [v_1,v_2]^\tsp $.
%
To model the intensity of undetected targets, a grid of size $ 201 \times 201 $ over $ \theta $ that covers the surveillance region is defined. This corresponds to grid cells of size $ 10\,\text{m} \times 10\,\text{m} $ centered around the points $ \theta^{(i)} = [p_1^{(i)},p_2^{(i)}]^\tsp $. 
The birth intensity is defined as in  \eqref{eq:ppp_prediction_approximation:birth_intensity} with
\begin{subequations}
\begin{align}
w^{\text{b}(i)}_k &= 
\begin{cases}
0.01/201, &i : p^{(i)}_1 = 1000,\\
0, &\text{otherwise,}
\end{cases} \\
\hat{\varphi} &= [-1,0]^\tsp\text{,} \\
P^{\varphi} &= I_2.
\end{align}
\end{subequations}
\ie, at each time step 0.01 new targets are expected to arrive from the right and travel towards the left with 1\,m$ / $s.

The mobile sensors move according to differential drive dynamics with constant speed of $ 5\, \text{m}/\text{s} $ and the control input, which is updated every ten seconds, determines the turn rate. 
Each of the admissible sequences of control inputs consists of a heading change performed with turn rate $ \pi/10\,\text{rad}/\text{s} $ followed by a straight path. The set of allowed heading changes for each sensor is $ \{n \pi / 6\,\text{rad},\ n \in \{-6,-5,\ldots,6\} \} $. The planning horizon is $ T=15 $ steps, \ie, effectively 150\,s.

\begin{figure*}
	\centering
	\includegraphics{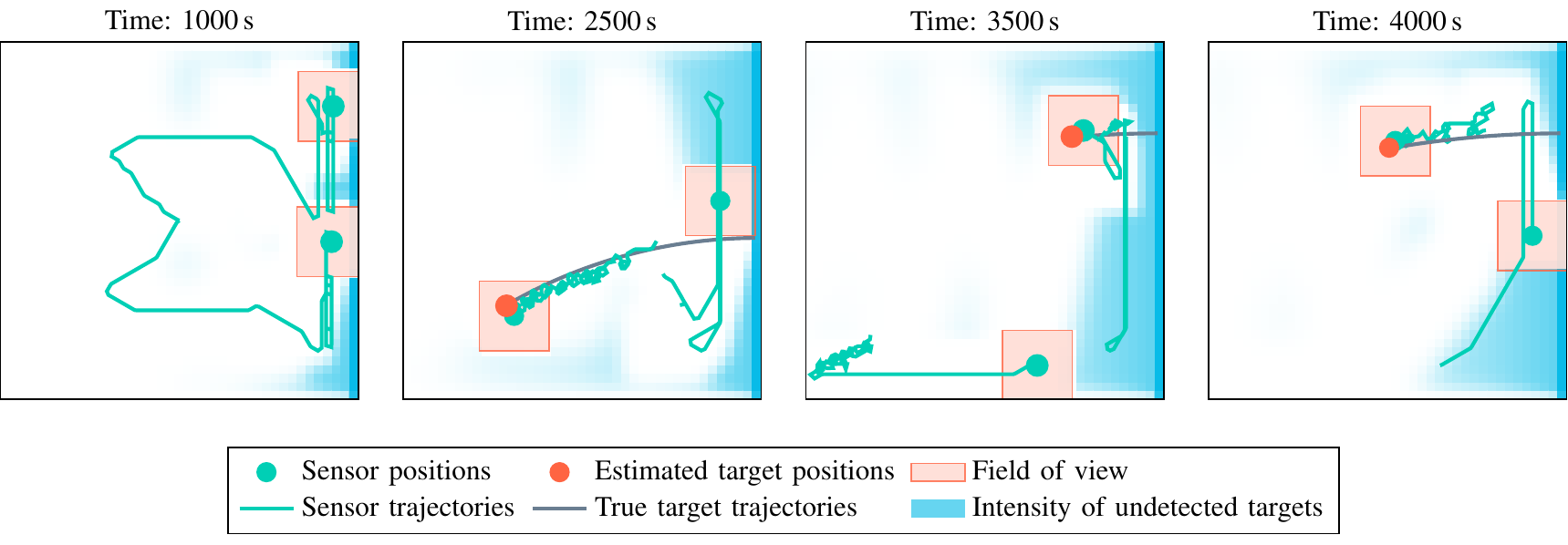}
	\caption{Snapshots of the simulated scenario where the sensor trajectories are planned for joint search and track based on the estimated PMBM density. The proposed representation of intensity of undetected targets is illustrated in white-blue scale, with a deeper blue indicating a higher intensity.}
	\label{fig:snapshots}
\end{figure*}
Fig.~\ref{fig:snapshots} shows snapshots from one realization of the scenario. New targets are expected to arrive from the right hand side, and the light blue areas illustrate the intensity of undetected targets. 
During the first 1000\,s of the simulation, when no targets are present, the two sensors patrol the right edge of the surveillance area in order to detect any appearing targets. When the first target appears, the two sensors split up. One of the sensors tracks the target and the other one continues to search for new targets along the edge of the area. As the second target arrives at time 3000\,s, the sensor that was patrolling the edge of the area has to track the new target while searching for undetected targets nearby. The first sensor continues to track the first target until it leaves the area, after which the sensor returns to the right edge of the area to search for new targets. The sensors' behavior indicates that the planning algorithm works as intended: the team utilizes the available resources to concurrently search for new targets and track the targets that have been detected.

To evaluate the tracking performance of the PMBM filter with the proposed grid-based birth intensity, it is compared to a conventional PMBM filter that uses a standard Gaussian mixture birth intensity. 
The Gaussian mixture birth intensity is designed to approximate the grid-based birth intensity using a reasonable number of components. It is given by ${ \lambda^\text{b} (x) = \sum_{i=1}^{N^\text{b}} w^{\text{b},i} \gaussian{x}{\hat{x}^{\text{b},i}}{P^{\text{b},i}} } $, with 
\begin{subequations}
	\begin{align}
		 N^\text{b} &= 9, \\
		 w^{\text{b},i}  &= 0.01 / N^\text{b}, \\
		 P^{\text{b},i}  &= \diag([1,1,125^2,1]), \\
		 \hat{x}^{\text{b},i}  &= [1000,-1,\hat{p}^{\text{b},i}_2,0]^\tsp \text{,}
	\end{align}
\end{subequations}
where $ \hat{p}^{\text{b},i}_2 = -1000 + 250(i-1) $ for $ i=1,\ldots,N^\text{b} $. 
The same scenario and planning algorithm parameters are used for both methods, the only difference is the underlying representation of the intensity of undetected targets.

\begin{figure}
	\centering
	\includegraphics{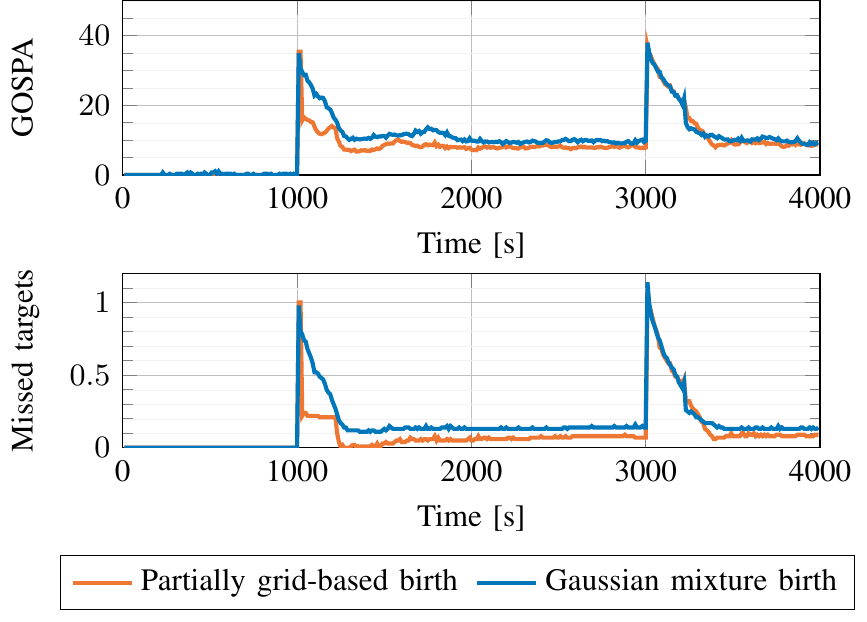}
	\caption{Comparison of performance measures for the proposed grid-based birth model and the conventional Gaussian mixture birth model.} 
	\label{fig:gospa}
\end{figure}

The generalized optimal subpattern assignment (GOSPA) metric~\cite{rahmathullah2017} is used to evaluate the performance of both methods. It is a unified performance metric for multi-target tracking that penalizes both localization errors for detected targets and errors due to missed and false targets.  
Fig.~\ref{fig:gospa} shows the GOSPA metric (with parameters $ \alpha=2 $, order two, localization error $ d(x,y) $ defined as the 2-norm of the position components of $ x-y $, and maximum allowable localization error 50\,m) averaged over 100 Monte Carlo runs. 	
The peaks in the GOSPA metric at 1000\,s and 3000\,s occur when new targets appear. The performance in terms of GOSPA is similar for both birth models. Fig.~\ref{fig:gospa} also shows that the sensors detect the targets more often, \ie, misses fewer targets, when the intensity of undetected targets is modeled using the grid-based model than when it is modeled using the Gaussian mixture.

\begin{figure}
	\centering
	\includegraphics{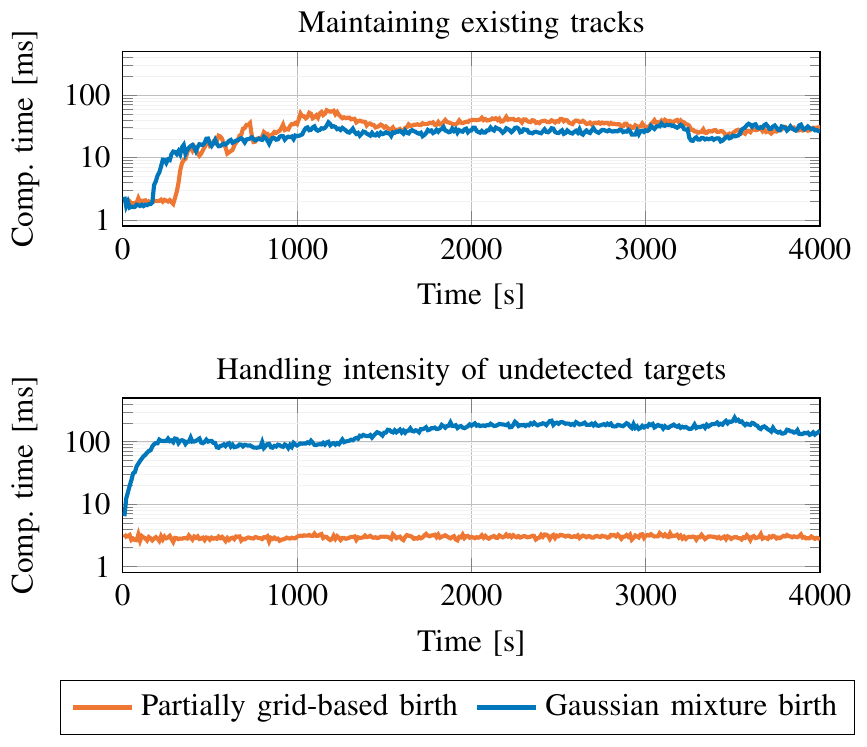}
	\caption{Computation time needed for the PMBM filter with the proposed grid-based birth model and with the conventional Gaussian mixture birth model.} 
	\label{fig:comptime}
\end{figure}

As a second performance measure, the computation time needed for a full filter iteration, \ie, a prediction step and a measurement update step, is analyzed. Although this comparison is based on rudimentary Matlab implementations, it gives an indication of the methods' relative computational complexity. For more insight, the computation time is split into two parts: (i) time needed for maintaining existing tracks, and (ii) time needed for handling the intensity of undetected targets and initiation of new tracks.
The computation times at each time step of the scenario averaged over 100 Monte Carlo simulations are shown in Fig.~\ref{fig:comptime}. 
The proposed method leads to slightly more time being spent on track maintenance. This is as expected, as fewer targets are missed with this method and the equations that correspond to track maintenance are the same regardless of the choice of birth model, see Appendix~\ref{sec:pmbm_filter_recursion}.
The computational benefit of the proposed method is a reduction in the time needed for handling the intensity of undetected targets and initiation of new tracks. 
For the partially grid-based birth model, the time spent on this step is fairly constant throughout the scenario and significantly less than the time spent on maintaining existing tracks.
This is in contrast to the Gaussian mixture birth model, where the time needed for this step depends on the number of components in the Gaussian mixture, which varies over time. As a large number of components are needed, significantly more time is spent on handling the intensity of undetected targets and initiation of new tracks than on maintaining existing tracks.


\section{Conclusions}
A PMBM filter with a new method to represent where yet undetected targets may be located has been derived. It relies on a grid-based intensity of undetected targets, which is useful in scenarios where the sensor's field of view does not cover the entire region of interest, as it allows for efficient representation of abrupt changes in the distribution. 
The intensity of undetected targets is estimated using Rao-Blackwellization, and the computational complexity of the prediction step is reduced using an approximation of an intermediate distribution.
The proposed method was compared to a conventional PMBM filter in a simulated sensor management application. 
The simulation study confirmed that the proposed method operates at significantly lower computational cost and provides similar tracking performance as the conventional filter.
The reduced computational complexity makes the proposed method suitable for sensor management applications, as these require methods that are both accurate and fast.

\bibliographystyle{IEEEtran}
\bibliography{IEEEabrv,refs}

\appendices
\section{Linear Gaussian PMBM filter recursion}\label{sec:pmbm_filter_recursion}
Assuming linear Gaussian dynamics and linear measurements according to
\begin{subequations}\label{eq:linagauss}
	\begin{align}
		p_{k+1,k}(x \cond x') &= \gaussian{x}{F x'}{Q} \label{eq:linagaussdyn} \text{,}\\
		p(z \cond x) &= \gaussian{z}{Hx}{R} \text{,} \label{eq:linagaussmeas}
	\end{align}
\end{subequations}
constant detection probability $ p_\textsc{d}(x) = p_\textsc{d} $, and constant survival probability $ p_\textsc{s}(x) = p_\textsc{s} $,
the prediction and update steps of the PMBM filter are given as follows.

\subsection{Prediction}
Let the posterior PMBM density be defined by the set of parameters
\begin{equation}\label{eq:pmbm_before_prediction}
\lambda^\text{u}_{k \mid k}, \{ w^j_{k \mid k}, \{ r^{j,i}_{k \mid k},p^{j,i}_{k \mid k} \}_{i \in \mathbb{I}^j_{k \mid k}}  \}_{j \in \mathbb{J}_{k \mid k}} \text{,}
\end{equation}
where the intensity of undetected targets $ \lambda^\text{u} $ is a Gaussian mixture
\begin{equation}
\lambda^\text{u}_{k \mid k} (x) = \sum_{i=1}^{N^\text{u}} w^{\text{u},i}_{k \mid k} \gaussian{x}{\hat{x}^{\text{u},i}_{k \mid k}}{P^{\text{u},i}_{k \mid k}} \text{,}
\end{equation}
and the spatial density of the \jth Bernoulli component in the \ith global hypothesis is Gaussian distributed according to $ p^{j,i}_{k \mid k}(x) = \gaussian{x}{\hat{x}^{j,i}_{k \mid k}}{P^{j,i}_{k \mid k}}  $. 
Then, with single-target transition model given by \eqref{eq:linagaussdyn} and Gaussian mixture birth intensity according to
\begin{equation}
\lambda^\text{b}_k (x) = \sum_{i=1}^{N^\text{b}} w^{\text{b},i}_k \gaussian{x}{\hat{x}^{\text{b},i}_k}{P^{\text{b},i}_k},
\end{equation}
the predicted PMBM density is on the same form as \eqref{eq:pmbm_before_prediction}.
The predicted intensity of undetected targets is given by
\begin{align}
\lambda^\text{u}_{k+1 \mid k} (x) &= 
\sum_{i=1}^{N^\text{b}} w^{\text{b},i}_{k+1} \gaussian{x}{\hat{x}^{\text{b},i}_{k+1}}{P^{\text{b},i}_{k+1}}
\nonumber \\
&+ \sum_{i=1}^{N^\text{u}} w^{\text{u},i}_{k+1 \mid k} \gaussian{x}{\hat{x}^{\text{u},i}_{k+1 \mid k}}{P^{\text{u},i}_{k+1 \mid k}},
\end{align}
where 
\begin{subequations}
	\begin{align}
		w^{\text{u},i}_{k+1 \mid k} &= 
		\pS 
		w^{\text{u},i}_{k \mid k} ,\\
		\hat{x}^{\text{u},i}_{k+1 \mid k} &= F \hat{x}^{\text{u},i}_{k \mid k} ,\\
		P^{\text{u},i}_{k+1 \mid k} &= F P^{\text{u},i}_{k \mid k} F^\tsp,
	\end{align}
\end{subequations}
the Bernoulli components are predicted according to
\begin{subequations}
	\begin{align}		
		r^{j,i}_{k+1 \mid k} &= 
		\pS
		r^{j,i}_{k \mid k} \text{,} \\
		\hat{x}^{j,i}_{k+1 \mid k} &= F \hat{x}^{j,i}_{k \mid k} \text{,}
		\\
		P^{j,i}_{k+1 \mid k} &= F P^{j,i}_{k \mid k} F^\tsp + Q\text{,}
	\end{align}
\end{subequations}
and $ 	w^j_{k+1 \mid k} = w^j_{k \mid k}  $, $ 	\mathbb{I}^j_{k+1 \mid k} = \mathbb{I}^j_{k \mid k}  $, $ 	\mathbb{J}_{k+1 \mid k} = \mathbb{J}_{k \mid k}  $.
%
%

\subsection{Measurement update}
\subsubsection{Data association}
As the true origins of measurements are unknown, association hypotheses are required. 
Let $ \mathbb{M} $ be an index set for the elements of the measurement set $ Z $, \ie,
\begin{equation}\label{eq:measurement_set}
Z = \{ z^m \}_{m \in \mathbb{M}}
\end{equation}
and let $ \mathcal{A}^j $ be a collection of all possible association hypotheses $ A $ for the $ j $th global hypothesis, \ie, the $ j $th MB, of which the targets are indexed by $ \mathbb{I}^j $. Then, an  association hypothesis $ A \in \mathcal{A}^j $ is a partition of $ \mathbb{M} \cup \mathbb{I}^j $ into nonempty disjoint subsets $ C \in  A $, called index cells \cite{granstrom2020}.

The standard assumptions in multi-target tracking that the targets are independent of each other implies that an index cell contains at most one target index and at most one measurement  index, \ie, $ \vert C\cap\mathbb{I}^j\vert \leq 1 $ and $ \vert C \cap \mathbb{M} \vert \leq 1 $ for all $ {C \in A \in \mathcal{A}^j} $. 
In the following, let $ i_C $ and $ m_C $ denote the target and measurement indices corresponding to index cell $ C $. 

\subsubsection{Update equations}
Let the predicted PMBM density be defined by the set of parameters
\begin{equation}\label{eq:pmbm_before_update}
\lambda^\text{u}_{k \mid k-1}, \{ w^j_{k \mid k-1}, \{ r^{j,i}_{k \mid k-1},p^{j,i}_{k \mid k-1} \}_{i \in \mathbb{I}^j_{k \mid k-1}}  \}_{j \in \mathbb{J}_{k \mid k-1}}
\end{equation}
where the intensity of undetected targets $ \lambda^\text{u} $ is a Gaussian mixture
\begin{equation}
\lambda^\text{u}_{k \mid k-1} (x) = \sum_{i=1}^{N^\text{u}} w^{\text{u},i}_{k \mid k-1} \gaussian{x}{\hat{x}^{\text{u},i}_{k \mid k-1}}{P^{\text{u},i}_{k \mid k-1}},
\end{equation}
and the spatial density of Bernoulli component $ j,i $ is Gaussian distributed according to $ p^{j,i}_{k \mid k-1}(x) = \gaussian{x}{\hat{x}^{j,i}_{k \mid k-1}}{P^{j,i}_{k \mid k-1}} $. Then, with a set of measurements $ Z $ and single-measurement likelihood according to \eqref{eq:linagaussmeas}, the updated PMBM density is a PMBM density given by
\begin{subequations}\label{eq:updated_pmbm}
	\begin{align}
	\pi^\textsc{pmbm}(X \cond Z) &= \sum_{X^\text{u} \uplus X^\text{d} = X} \pi^\textsc{p} (X^\text{u}) \pi^\textsc{mbm}(X^\text{d}), \\\
	\pi^\textsc{p}(X^\text{u}) &= e^{- \langle \lambda^\text{u}_{k \mid k},1 \rangle}  \prod_{x \in X^\text{u}} \lambda^\text{u}_{k \mid k}(x), \\
	\pi^\textsc{mbm}(X^\text{d}) &= \sum_{j \in \mathbb{J}_{k \mid k-1}} \sum_{A \in \mathcal{A}^j} w^j_A \pi^j_A(X^\text{d}),\\
	\pi^j_A(X^\text{d}) &= 
	\sum_{\uplus_{C \in A} X^{i_C} = X^\text{d}}  \prod\limits_{C \in A} \pi^{j}_C (X^{i_C}),
	\end{align}
\end{subequations}
where the weights of the global hypotheses are given by
\begin{equation}\label{eq:updated_hypweights}
	w_A^j = \frac
	{w^j_{k \mid k-1} \prod_{C \in A} \mathcal{L}_C}
	{\sum_{j \in \mathbb{J}_{k \mid k-1}} \sum_{A \in \mathcal{A}^j} w^j_{k \mid k-1} \prod_{C \in A} \mathcal{L}_C} 
\end{equation}
and the parameters of \eqref{eq:updated_pmbm} and \eqref{eq:updated_hypweights} are described in the following.

\paragraph{Intensity of undetected targets}
The updated intensity of undetected targets $ \lambda^\text{u}_{k \mid k} $ is a Gaussian mixture, with weights updated according to
\begin{equation}
w^{\text{u},i}_{k \mid k} = 
p_\textsc{d}
w^{\text{u},i}_{k \mid k-1}
\end{equation}
and unchanged mixture components, \ie, $ 	\hat{x}^{\text{u},i}_{k \mid k} = \hat{x}^{\text{u},i}_{k \mid k-1} $ and $
P^{\text{u},i}_{k \mid k} = P^{\text{u},i}_{k \mid k-1} $.

\paragraph{Potential target detected for the first time}
Each measurement $ z_k^m $ generates a new Bernoulli component with existence probability and density according to
\begin{subequations}
	\begin{align}
	r_{k \mid k} &= \frac{ e_k (z_k^m)}
	{\lambda^\text{fa}(z^{m}) +  e_k (z_k^m) } \\
	p(x) &= \frac{1}{e_k(z_k^m)} \sum_{i=1}^{N^\text{u}} c_k^{\text{u},i} \gaussian{x}{\hat{x}_{k \mid k}^{\text{u},i}}{P_{k \mid k}^{\text{u},i}} \label{eq:new_bernoulli_density}
	\end{align}
\end{subequations}
where
\begin{subequations}
	\begin{align}
	\hat{x}_{k \mid k}^{\text{u},i} &= \hat{x}_{k \mid k-1}^{\text{u},i} + K_k (z_k^m - H \hat{x}_{k \mid k-1}^{\text{u},i})\\
	P_{k \mid k}^{\text{u},i} &= P_{k \mid k-1}^{\text{u},i} - K_k H P_{k \mid k-1}^{\text{u},i} \\
	K_k &= P_{k\mid k-1}^{\text{u},i} H^\tsp (S_k^{\text{u},i})^{-1} \\
	S_k^{\text{u},i} &= H P_{k\mid k-1}^{\text{u},i} H + R  \\
	c_k^{\text{u},i} &= 
	p_\textsc{d}
	w_{k \mid k-1}^{\text{u},i}
	\gaussian{z_k^m}{H \hat{x}_{k \mid k-1}^{\text{u},i}}{S_k^{\text{u},i}} \\
	e_k (z_k^m) &= \sum_{i=1}^{N^\text{u}} c_k^{\text{u},i}
	\end{align}
\end{subequations}
The Gaussian mixture in \eqref{eq:new_bernoulli_density} is  approximated as a Gaussian distribution by performing moment matching.
The likelihood corresponding to the hypothesis that measurement $ z_k^m $ originates from a previously undetected target is given by
\begin{equation}
\mathcal{L}_C = \lambda^\text{fa}(z^{m}) + e_k(z_k^m)
\end{equation}
with $ C\cap\mathbb{I}^j_{k \mid k-1} = \emptyset $ and $  C \cap \mathbb{M}  = m $.

\paragraph{Previously detected targets}
A predicted Bernoulli component with existence probability $ r_{k \mid k-1}^{j,i} $ and spatial density $ p^{j,i}_{k \mid k-1}(x) = \gaussian{x}{\hat{x}_{k \mid k-1}^{j,i}}{P_{k \mid k-1}^{j,i}} $ generates a misdetection hypothesis and one hypothesis for each measurement in $ Z $.

Under the hypothesis that the target is misdetected, the updated existence probability is given by
\begin{equation}
r_{k \mid k} =  \frac{r_{k \mid k-1}^{j,i}(1 -
p_\textsc{d}
)}{1-r_{k \mid k-1}^{j,i} + r_{k \mid k-1}^{j,i}(1 - 
p_\textsc{d}
)}
\end{equation} 
and the density remains the same, \ie, $ 	\hat{x}_{k \mid k}^{j,i} = \hat{x}_{k \mid k-1}^{j,i} $ and $ P_{k \mid k}^{j,i} = P_{k \mid k-1}^{j,i} $.
The likelihood corresponding to this hypothesis is given by 
\begin{equation}
\mathcal{L}_C = 1-r_{k \mid k-1}^{j,i} + r_{k \mid k-1}^{j,i}(1 - 
p_\textsc{d}
)
\end{equation}
with $ C\cap\mathbb{I}^j_{k \mid k-1} = i $ and $  C \cap \mathbb{M}  = \emptyset $.

Under the hypothesis that the target is detected with measurement $ z_k^m $, the Bernoulli component has existence probability $ r_{k \mid k} = 1 $ and density $ \gaussian{x}{\hat{x}_{k \mid k}^{j,i}}{P_{k \mid k}^{j,i}} $ where
\begin{subequations}
	\begin{align}
	\hat{x}_{k \mid k}^{j,i} &= \hat{x}_{k \mid k-1}^{j,i} + K_k (z_k^m - H \hat{x}_{k \mid k-1}^{j,i})\\
	P_{k \mid k}^{j,i} &= P_{k \mid k-1}^{j,i} - K_k H P_{k \mid k-1}^{j,i} \\
	K_k &= P_{k\mid k-1}^{j,i} H^\tsp (S_k^{j,i,m})^{-1} \\
	S_k^{j,i,m} &= H P_{k\mid k-1}^{j,i} H + R 
	\end{align}
\end{subequations}
The likelihood corresponding to this hypothesis is given by 
\begin{equation}
\mathcal{L}_C = r_{k \mid k-1}^{j,i} 
p_\textsc{d}
\gaussian{z^{m}}{H \hat{x}^{j,i}_{k \mid k-1}}{S^{j,i,m}_{k \mid k-1}}
\end{equation}
with $ C\cap\mathbb{I}^j_{k \mid k-1} = i $ and $  C \cap \mathbb{M}  = m $.

\end{document}